\documentclass[%
 reprint,
 amsmath,amssymb,
 aps,
 prr
]{revtex4-2}

\usepackage{graphicx}
\usepackage{dcolumn}
\usepackage{bm}

 \usepackage{xcolor}
\usepackage{dirtytalk}
\usepackage{graphicx}
\usepackage{caption}
\usepackage{subcaption}
\usepackage{soul}

 \def\<#1>{\mathinner{\langle#1\rangle}}

	\newcommand*\diff{\mathop{}\!\mathrm{d}}
 
	\newcommand*\eff{\mathop{}\!\mathrm{eff}}

\begin{document}

\preprint{APS/123-QED}

\title{A compact stellarator-tokamak hybrid}

\author{S.A. Henneberg}%
 \email{Sophia.Henneberg@ipp.mpg.de}
\affiliation{
Max-Planck-Institut für Plasmaphysik, Wendelsteinstr. 1, 17489 Greifswald
}%
\author{G.G. Plunk}%
\affiliation{
Max-Planck-Institut für Plasmaphysik, Wendelsteinstr. 1, 17489 Greifswald
}%

\date{\today}

\begin{abstract}
Tokamaks and stellarators are the leading two magnetic confinement devices for producing fusion energy, begging the question of whether the strengths of the two could be merged into a single concept. To meet this challenge, we propose a first-of-its kind optimized stellarator-tokamak hybrid. Compared to a typical tokamak coil set, only a single simple type of stellarator coil has to be added which leads to a compact, volume- and transport-preserving magnetic field, with an added rotational transform that reaches levels thought to enhance stability.
\end{abstract}

\maketitle


Tokamaks and stellarators have comparative weaknesses:  Stellarators are often criticized for the relatively small plasma volume they achieve, and for their complicated electromagnetic coils, both numerous in type and difficult to build.  On the other hand, tokamaks rely on plasma currents to generate the magnetic field, which can generate detrimental instabilities and which impedes a desired steady-state operation.

The idea of a stellarator-tokamak hybrid is simple and compelling: to combine the strengths of the two concepts into a single device.  Ideally, it would offer large plasma volume (compactness), easily built coils, and simple and inherently steady-state operation.  However, this has proved an elusive combination to realize.

Several stellarator-tokamaks hybrids designs have been proposed in the past, such as the  spherical stellarator concept \cite{Moroz-1996, Moroz-1998} or the tokastar \cite{Yamazaki-1985}. Hybrid machines like W7-A \cite{W-VII-A-Team-1980,Hirsch-2008} and the Compact Toroidal Hybrid (CTH) device \cite{Pandya-2015,Hartwell-2017} have even been built and operated, yielding valuable insight into how the three-dimensional shaping of tokamaks can enhance stability. However, none of these hybrids have persevered transport properties, as stellarators have, by default, higher so-called neoclassical transport, and only carefully tailored stellartors perform as well as tokamaks in this regard \cite{Landreman-2022-a}.

Any hybrid device that could be considered as a reasonable basis for fusion energy must therefore be optimized for neoclassical transport.  This suggests that the optimized class most closely related to tokamaks, so-called quasi-axisymmetric (QA) stellarators \cite{Nuehrenberg-1994,Helander-2014-a}, must be considered.  For these stellarators, the magnetic field strength possesses a hidden symmetry that is revealed upon transformation to special (``Boozer'') coordinates \cite{Boozer-1981-a}.

QA stellarators are thus the natural choice for designing a hybrid, but existing designs possess a highly twisted shape that departs strongly from the simple toroidal symmetry of a tokamak.  This allows such devices to generate significant rotation of the magnetic field lines, imparting confinement and the passive stability that characterizes stellarators.  However, a quick survey of previous QA designs, for instance \cite{Nuehrenberg-1994,Zarnstorff-2001,Drevlak-2013,Henneberg-2019,Landreman-2022-a}, reveals a downside of this strong shaping, namely that the axisymmetric volume contained within the stellarator coils is drastically smaller than that available for stellarator operation. This means the tokamak part in such a flexible hybrid would have an aspect ratio of above 5. A large portion of the volume of the device, one of the main contributors to the cost, would therefore be wasted.  It is hard to conceive of a tokamak-like hybrid device based on such designs.

\begin{figure}
    \centering
     \includegraphics[trim= 8cm 0cm 0cm 0cm,clip,width=.99\linewidth] {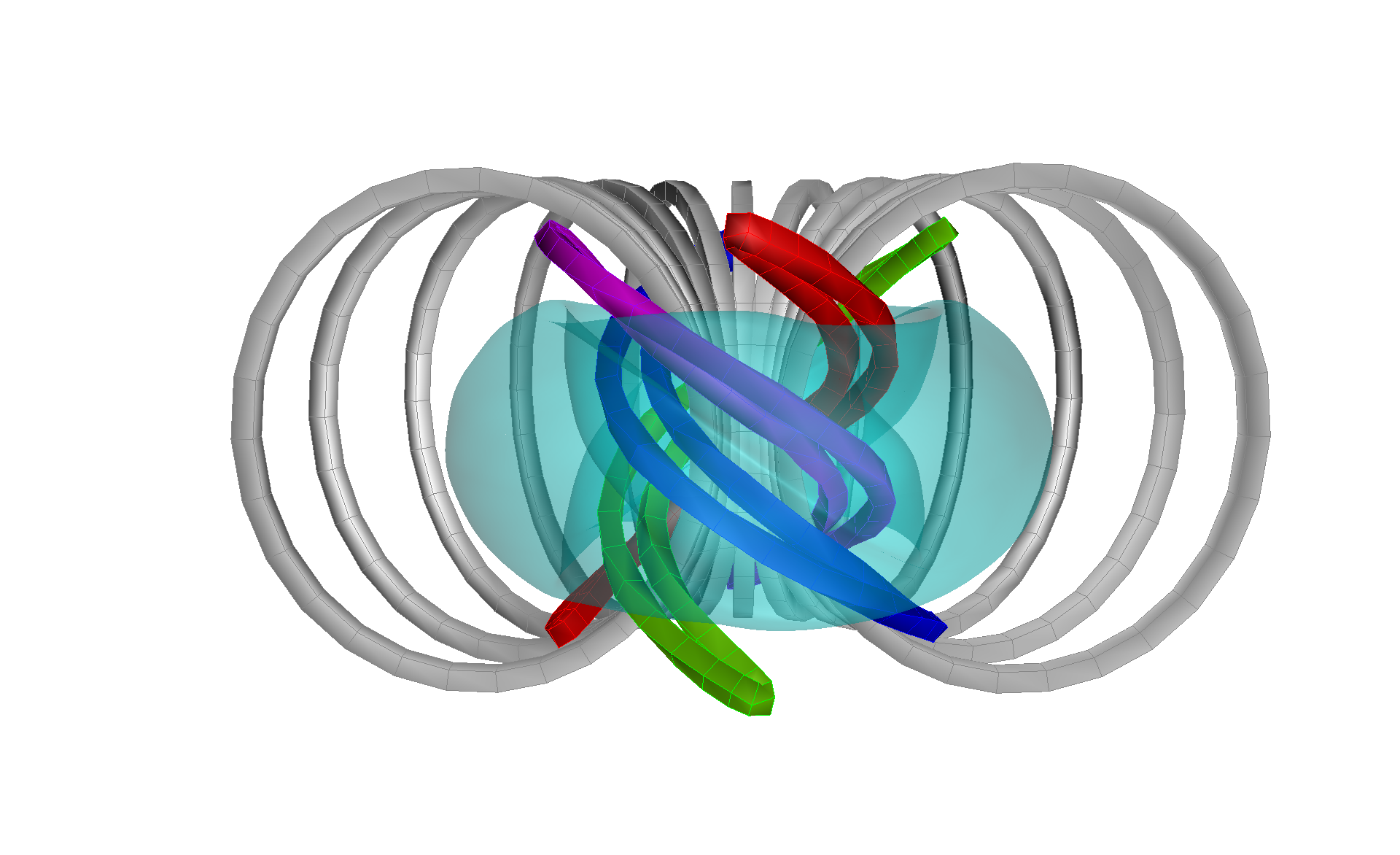}
    \caption{The plasma boundary (\textcolor{cyan}{cyan}) encompassing four identical QA (``banana") coils (\textcolor{red}{red}, \textcolor{green}{green}, \textcolor{blue}{blue}, and \textcolor{violet}{purple}) all contained within toroidal field coils (\textcolor{gray}{gray}; half of them shown).}
    \label{fig:full_coil_set}
\end{figure}

\begin{figure*}
\centering
\includegraphics[trim= 0.1cm 0.5cm 2cm 0.1cm,clip,width=18.cm]{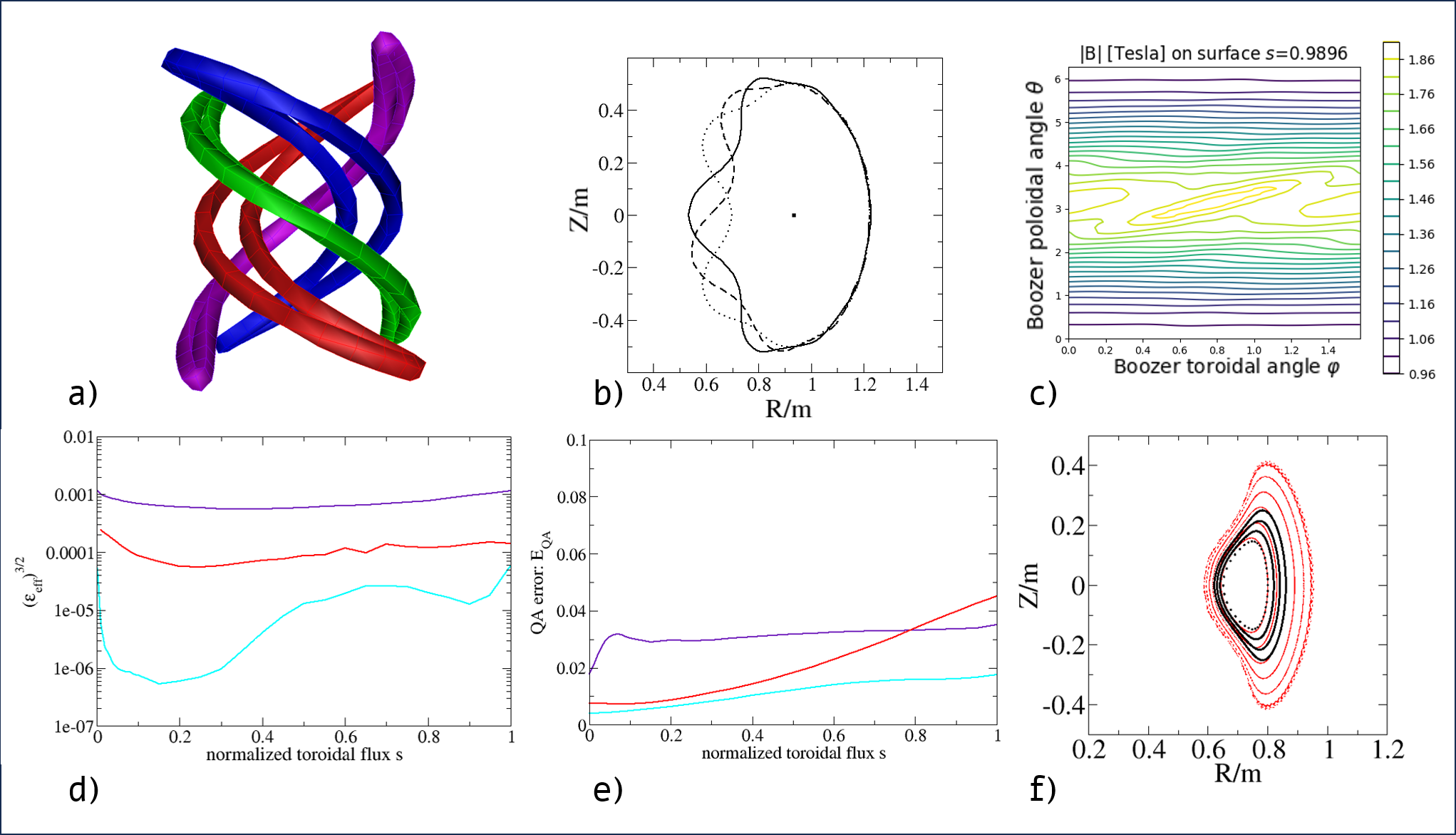}
\caption{Overview of results for the hybrid coil set: a) The four identical QA (``banana") coils -- the only coils needed in addition to the standard tokamak coils. b) The cross-sections of the finite beta plasma boundary with the original profiles evaluated with free-boundary VMEC \cite{Hirshman-1983} at three toroidal locations ($\varphi= 0^\circ, 22.5^\circ, 45^\circ$). c) The contour lines of magnetic field strength with respect to Boozer angles, \cite{Boozer-1981-a}, near the plasma boundary for the original plasma profiles. d) The effective ripple \cite{Nemov-1999} to the power of $3/2$ and e) the quasi-axisymmetric error, given by Eq.~\ref{eq:QAerror}, (both evaluated based on VMEC-outputs) vs the normalized toroidal flux $s$ for three different profiles presented in this paper: the original profiles (\textcolor{cyan}{cyan}), the altered profiles from \cite{Henneberg-2019} (\textcolor{red}{red}) and the equilibrium with zero beta and zero net toroidal plasma current (\textcolor{violet}{purple}). e) Poincar\'{e} plots showing the flux surfaces in vacuum produced by solely TF and QA coils with the original coil currents in \textcolor{black}{black} and with altered QA coil current to increase volume in \textcolor{red}{red} at a single toroidal location, $\varphi=0^\circ$ for clarity. }\label{fig:combined}
\end{figure*}

Instead, we devise a hybrid concept, shown in Fig.~\ref{fig:full_coil_set}, based on a class of QA solutions found as perturbations of tokamak equilibria \cite{Plunk-2018, Plunk-2020}, which achieve significant rotational transform from non-axisymmetric shaping without significantly disturbing the overall plasma volume. The rotational transform, measuring the number of poloidal turns a field line makes per toroidal turn \cite{Helander-2014-a}, is known to be critical to stability and confinement. These special perturbed tokamak solutions are found by venturing into an unfamiliar area of design space (lower aspect ratio and/or higher field periods) that so far has seemed inaccessible to QA stellarators \cite{Landreman-2022}.  In this limit, the external rotational transform is intuitively generated by localized stellarator shaping on the inboard of the torus, in the form of a tightly grooved pattern in the magnetic surfaces; see Fig. \ref{fig:combined}(b).  This odd arrangement seems to be the key to realizing a tokamak-like equilibrium with the character ({\em e.g.} external rotational transform) of a stellarator, and is the basis of our idea for a tokamak-stellarator hybrid.

\begin{figure}
    \centering
    \includegraphics[width=0.43\textwidth]{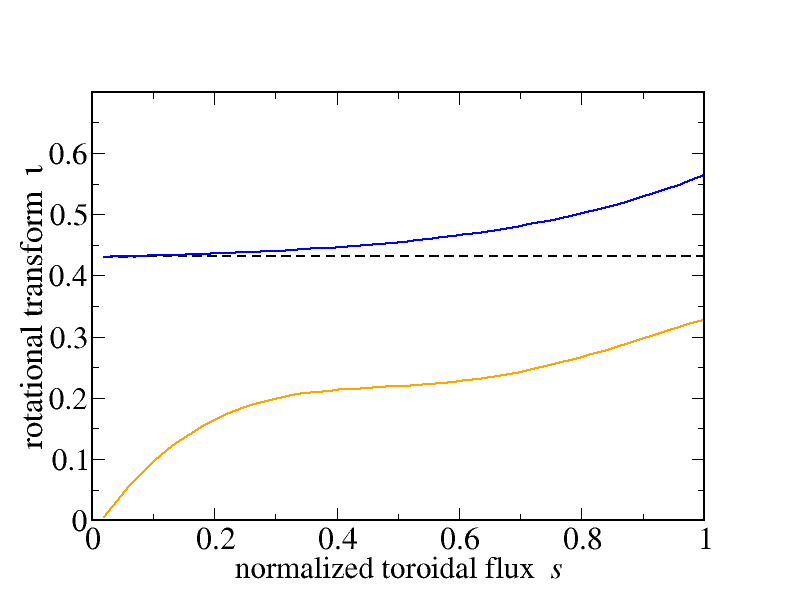}
    \caption{Rotational transform, $\iota$, versus normalized flux $s$; The unperturbed, targeted case is in dashed \textcolor{black}{black}, the perturbed, targeted case is in solid \textcolor{blue}{blue}, the rotational transform with altered profiles \cite{Henneberg-2019} (see Fig. \ref{fig:PlasmaProfiles}) in solid \textcolor{orange}{orange}.}
    \label{fig:rotational_transform}
\end{figure}

\begin{figure}
\centering
\begin{subfigure}{.48\columnwidth}
  \centering
  \includegraphics[trim= 1cm 1cm 4cm 1cm,clip,width=1.0\linewidth]{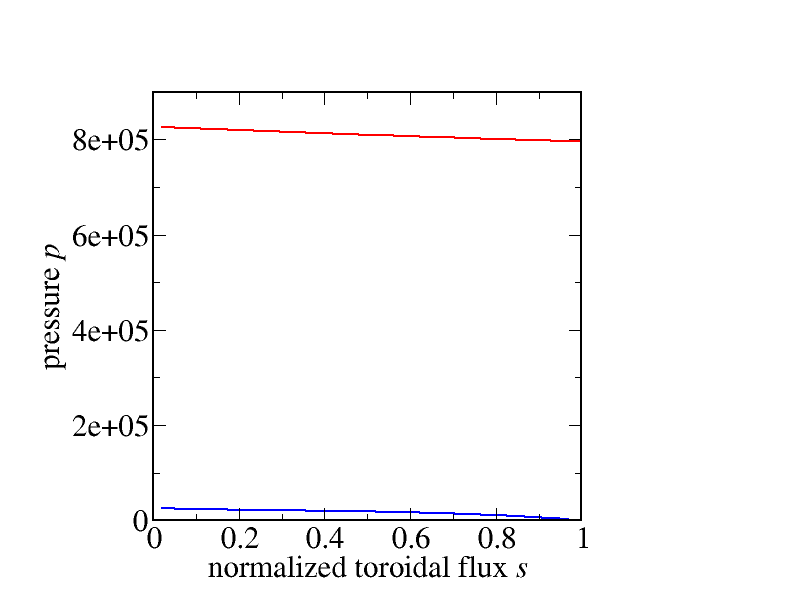}
  \caption{}
\end{subfigure}
\begin{subfigure}{.48\columnwidth}
  \centering
  \includegraphics[trim= 0cm 1.cm 5cm 2cm,clip,width=1.\linewidth, ]{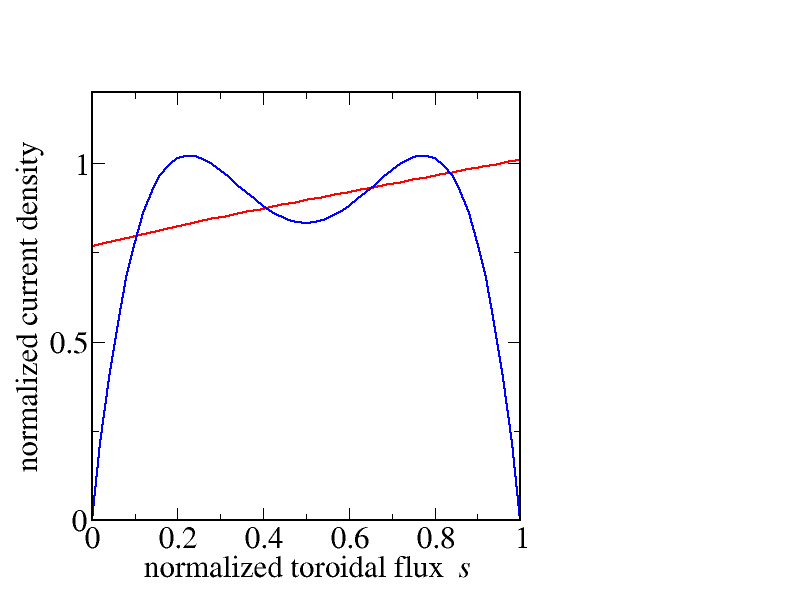}
 \caption{}
  \label{fig:side_view_tok_qa_RMPs}
\end{subfigure}
\caption{The plasma profiles used in the analytic work and the evaluation of the coils based on \cite{Plunk-2020} in \textcolor{red}{red} as well as the altered plasma profiles based on \cite{Henneberg-2019} in \textcolor{blue}{blue}. Left: pressure profiles vs normalized toroidal flux. Right: normalized current density vs normalized toroidal flux. }
\label{fig:PlasmaProfiles}
\end{figure}

The perturbed axisymmetric equilibrium, which we investigate in this paper, has a field period number of 4, meaning that it consists of 4 identical torus segments, and it has a net toroidal plasma current $I_{tor}$ of $\approx 574$kA. The original tokamak has a flat rotational transform profile of around $0.43$, while the rotational transform of the original perturbed axisymmetric equilibrium ranges from $0.43$ at the axis an increases to $0.56$ at the plasma boundary, Fig.~\ref{fig:rotational_transform}. The pressure and current density profiles are the ones used in \cite{Plunk-2020} and are shown in Fig. \ref{fig:PlasmaProfiles} for convenience.

For the coils presented in this paper, we performed the second stage of the traditional two-stage approach \cite{Henneberg-2021}. This means that we optimized the coil geometry such that they can reproduce the plasma boundary found from the analytic theory by minimizing the normal components of the magnetic field on the fixed plasma boundary given by the theory. For this approach the profiles within the plasma are also fixed. Note that we did not optimize the plasma boundary any further, though it should certainly be possible to improve upon the perturbative near-axisymmetric solution. We will leave this task for future work.

For the coil optimization, we use the coil optimization features of the optimization framework simsopt~\cite{Landreman-2021,simsopt-version-2022}. 
We target, as typical in the two-stage approach, the quadratic-flux error functional:
\begin{equation}
   \varphi_2= \frac{1}{2}\int |B_{E,n}-D_n|^2 \diff s, 
\end{equation}
where $\int \diff s$ is a surface integral over the plasma boundary, $B_{E,n}$ is the normal component of the external magnetic field ${\bf B}_E$ (produced by the coils) with respect to the plasma boundary. $D_n$ is the targeted normal component of the external magnetic field. In vacuum $D_n=0$, since the plasma boundary is a flux surface. When investigating finite-beta plasmas, as done here, one has to determine the magnetic field produced by plasma currents to be able to calculate $D_n$. We used the so-called virtual casing method \cite{Shafranov-1972} for this purpose. The virtual casing methods provides the plasma current contribution to the normal component of the magnetic field strength on the plasma boundary by only evaluating a surface integral. For that one only needs the tangential part of the magnetic field which is determined with a fixed-boundary VMEC calculation \cite{Hirshman-1983}.
In addition, we target the length of the coils $L=\sum_j L_j $ with a scalar penalty functional $Q\equiv \varphi_2 + (L_t-L)^2$,
where $L_t$ is a user specified target value for the sum of the compound coil length. Note that we choose a particularly simple target function with only two terms to focus on finding suitable coils, and plan to refine the target in future work.

Axisymmetric equilibria require a net toroidal plasma current to produce rotational transform and flux surfaces, inducing a significant vertical component in the targeted normal component of the external magnetic field on the plasma boundary, $D_n$. To be able to obtain the typical tokamak toroidal field (TF) coils in the optimization procedure, one has to add poloidal field (PF) coils to cancel the vertical component of $D_n$. 
We started with four PF coils, where we only allowed the radius and the height $z$ to vary. \\
Next we added 20 TF coils, with a large radius chosen to minimize coil ripple effects, as the geometry of the additional QA coils is our main interest; smaller and/or fewer TF coils may be used in future studies.  These coils need not be optimized since only their total current is determined by the given equilibrium.

The perturbed axisymmetric equilibrium chosen for this study has four field periods and is stellarator symmetric. During optimization we initialized the additional QA coils as circular shapes located at the inboard side (not interlinked with the plasma boundary); the TF and PF coils geometries were held fixed during this optimization. Conventionally, stellarators employ so-called modular coil designs, like in Wendelstein 7-X or NCSX. However, to realize the unusual shape of our hybrid configuration, we found it more successful to choose  QA coils of the that are not interlinked with the vacuum vessel, similar to saddle coils, except more elongated and carefully shaped.

To evaluate the quality of our quasi-symmetry we use the quasi-symmetric error defined as:
\begin{equation}
E_{QA}=\left(\sqrt{\sum_{n\neq0,m}B_{m,n}^2}\right)/B_{00} \label{eq:QAerror}
\end{equation}
where the $B_{mn}$ are the Fourier coefficients of the magnetic field in Boozer coordinates. Another measure used is the effective ripple $\epsilon_{\eff}$, which is a proxy for the neoclassical transport~\cite{Nemov-1999}. To aid comparison we present the effective ripple to the power $3/2$.

Using free-boundary version of the ideal MHD equilibrium solver VMEC \cite{Hirshman-1983} with the magnetic field produced by our coils, we obtain the new plasma boundary and flux surfaces. To test if the quasi-axisymmetry properties are maintained with the realization with the coils, we determine the quasi-axisymmetric error from these VMEC equilibria. 

 Since it is not guaranteed that flux surfaces exists in vacuum~\cite{Helander-2014-a}, we evaluate the existence, shape, and volume of the flux surfaces by generating Poincar\'{e} plots based only on the coils' magnetic field. We used the ROSE/ONSET suites~\cite{Drevlak-2019} to generate the Poincar\'{e} plots, and the plots of the quasi-axisymmetric error and the effective ripple.

\textit{Results and Discussions --}
In addition to a set of standard tokamak  coils (toroidal field (TF) and poloidal field (PF) coils \cite{Wesson-2004}), we find that a single type of simple ``QA coil'' on the inboard side is sufficient to reproduce the perturbed QA equilibrium; see Fig. \ref{fig:combined} a).

The equilibrium determined by the free-boundary VMEC code recreates the key feature of the boundary found from the analytic theory: the perturbation of the tokamak is dominantly located on the inboard side and the outboard side is mostly unaltered, see Figs. \ref{fig:combined} b). 
The contours of constant magnetic field strength, plotted versus Boozer angles, look nearly straight everywhere except near the inboard mid-plane, and the Fourier harmonics also confirm the quasisymmetry feature, Fig. \ref{fig:combined} c).

Both the effective ripple and the quasi-axisymmetric error are measures of how well the stellarator is transport optimized, e.g. the effective ripple of CHS or TJ-II are as high as 30\%-40\% \cite{Beidler-2011}. The effective ripple, Fig.~\ref{fig:combined} d), and the quasi-axisymmetric errors, Fig.~\ref{fig:combined} e), are small when evaluated with the original profiles but also, surprisingly, with altered plasma profiles Fig.~\ref{fig:PlasmaProfiles}. For the altered plasma profiles we choose the one from \cite{Henneberg-2019} and the case of zero net toroidal current and zero plasma beta where the plasma beta $\beta$ is the ratio of plasma pressure $p$ and magnetic pressure $B^2/(2\mu_0)$. Even in this extreme case the effective ripple is below one percent nearly everywhere which is comparable or even lower than in the largest transport optimized stellarator Wendelstein 7-X or in the quasi-axisymmetric design NCSX \cite{Beidler-2011}.  

To further evaluate quasi-symmetry quality, we calculated fast particle losses of fusion born alpha particles (3.5MeV), initialized at half radius ({\em e.g.} quarter flux $s=0.25$).  Only 1.45\% were lost after 0.2sec. This was evaluated with the SIMPLE code where 5000 particles were isotropically launched and followed, assuming no collisions \cite{Albert-2020,Albert-2020-b}. The size of the machine was scaled to have the same minor radius of the QA reactor design ARIES-CS \cite{Najmbadi-2008}. The observed confinement is better than most historic QA designs \cite{Landreman-2022-a}, though we emphasize that no QA optimization was performed (beyond the use of the approximately QA analytic solution) and also note that we use a magnetic field produced by coils, which typically have worse confinement due to field errors. 

We find clear evidence of vacuum flux surfaces in the Poincar\'{e} plots shwon in Fig. \ref{fig:combined} f). 
In such an experiment one has the freedom to control the volume of these flux surfaces by changing the QA coil current (see Fig. \ref{fig:combined} f) and also to control the added external rotational transform, reaching levels anywhere from zero to 0.3. 
This is in the range that has been shown in experiments to improve stability properties \cite{W-VII-A-Team-1980,Hirsch-2008,Pandya-2015,Hartwell-2017}. In these experiments it was found that an external rotational transform of around 0.1 to 0.15 can suppress disruptions.

A machine based on such a design would be a suitable candidate to study how 3D shaping affects plasma properties, such as stability, while approximately maintaining small neoclassical transport. In addition, one would be able to investigate if the plasma current can be ramped up starting from the vacuum flux surfaces generated by the quasi-axisymmetric coil currents. This ``quasi-axisymmetric start-up'' could be an alternative to ordinary Ohmic tokamak start-up scenarios. Thus, there would be no need for a central solenoid, which is one of the most complex and expensive components of a conventional tokamak.

Since the QA coils are only on the inboard side, it might be possible to simply ``upgrade" an existing tokamak by adding such coils. We note that inboard-type coils have indeed been realized in the past, e.g. at TEXTOR as part of a divertor concept \cite{Neubauer-2005}. However, the coils in this work will be larger than in TEXTOR and it might be necessary to remove the solenoid. In this case, our proposed alternative QA-startup could be used.

For a major radius of around 1m as we have presented here, the minimum coils-qa-plasma distance is 15cm but only 6cm for the distance between the QA-coils and the axisymmetric plasma boundary. If one scales this machine to a reactor size, e.g. of 1900m$^3$, this shortest coil-to-plasma distance increases to approximately 1.5m. This distance for the QA operation is already (without including the distance into the optimization) relevant for a reactor-sized device, for which it is said that at least 1.5m of space is needed for a breading blanket and neutron shielding.

The aspect ratio of the QA stellarator depends on the chosen axisymmetric equilibrium which is perturbed with the analytic model outlined in \cite{Plunk-2018,Plunk-2020}. This suggests that at any aspect ratio a hybrid design could be generated with the approach presented here as long as the QA coils can be made to fit in the center. Therefore our approach should permit designs comparable in compactness with previous tokamak-stellarator hybrids including the so-called ``spherical stellarator"~\cite{Moroz-1996}, while including the additional benefits described here, {\it i.e.} transport optimization, coil simplicity, {\it etc}. 

The work presented here is an initial step in exploring our compact tokamak-hybrid concept.  The equilibrium solutions so far considered represent a small part of the available space to explore.  Indeed, any viable tokamak design can be considered a candidate for the application of our method (perhaps even axisymmetric equilibria that might not be considered for pure tokamak operation).  This opens up a large optimization space for possible hybrid designs, over which quasi-axisymmetry and other properties can be directly targeted, and further improved.  MHD stability, divertor designs and even micro-turbulence should also be investigated to see if the stellarator and tokamak strengths might also be favorably combined.  These are just some of the possible activities to expand and refine our concept of the stellarator-tokamak hybrid.

\begin{acknowledgments}
We would like to thank Per Helander, Michael Drevlak, Robert Davies, and Brendan Shanahan for helpful discussions, Samuel Lazerson for the support with XGRID as well as helpful discussions, Caoxiang Zhu for helping converting coil types within simsopt, Matt Landreman with general support with simsopt, Joachim Geiger for his support using free-boundary VMEC, and the simsopt development team.
This work has been carried out within the framework of the EUROfusion Consortium, funded by the European Union via the Euratom Research and Training Programme (Grant Agreement No 101052200 — EUROfusion). Views and opinions expressed are however those of the author(s) only and do not necessarily reflect those of the European Union or the European Commission. Neither the European Union nor the European Commission can be held responsible for them.
\end{acknowledgments}

\bibliography{apssamp}

\end{document}